# Interval edge-colorings of graph products

Petros Petrosyan

Institute for Informatics and Automation Problems of NAS of RA,
Department of Informatics and Applied Mathematics, YSU,
Yerevan, Armenia
e-mail:
pet_petros@ipia.sci.am

Hrant Khachatrian,
Liana Yepremyan

Department of Informatics and Applied Mathematics, YSU,
Yerevan, Armenia
e-mails: hrant@egern.net,
liana.yepremyan2009@gmail.com

Hovhannes Tananyan

Department of Applied Mathematics and Informatics, RAU,
Yerevan, Armenia
e-mail:
HTananyan@yahoo.com

## ABSTRACT

An interval $t-$coloring of a graph $G$ is a proper edge-coloring of $G$ with colors $1, 2, \ldots, t$ such that at least one edge of $G$ is colored by $i$, $i = 1, 2, \ldots, t$, and the edges incident to each vertex $v \in V(G)$ are colored by $d_G(v)$ consecutive colors, where $d_G(v)$ is the degree of the vertex $v$ in $G$. In this paper interval edge-colorings of various graph products are investigated.

## Keywords
Edge-coloring, interval coloring, products of graphs, regular graph

## 1. INTRODUCTION

All graphs considered in this paper are finite, undirected, connected and have no loops or multiple edges. Let $V(G)$ and $E(G)$ denote the sets of vertices and edges of a graph $G$, respectively. The degree of a vertex $v \in V(G)$ is denoted by $d_G(v)$, the maximum degree of a vertex in $G$ by $\Delta(G)$, the chromatic index of $G$ by $\chi'(G)$, and the diameter of $G$ by $d(G)$. We use the standard notations $P_n, C_n, K_n$ and $Q_n$ for the simple path, simple cycle, complete graph on $n$ vertices and the $n-$dimensional cube, respectively.

Let $G$ and $H$ be two graphs.
The Cartesian product $G \square H$ is defined as follows:
$$V(G \square H) = V(G) \times V(H),$$
$$E(G \square H) = \{(u_1, v_1)(u_2, v_2) \mid u_1 = u_2 \text{ and } v_1 v_2 \in E(H) \text{ or } v_1 = v_2 \text{ and } u_1 u_2 \in E(G)\}.$$

The tensor (direct) product $G \times H$ is defined as follows:
$$V(G \times H) = V(G) \times V(H),$$
$$E(G \times H) = \{(u_1, v_1)(u_2, v_2) \mid u_1 u_2 \in E(G) \text{ and } v_1 v_2 \in E(H)\}.$$

The strong tensor (semistrong) product $G \otimes H$ is defined as follows:
$$V(G \otimes H) = V(G) \times V(H),$$
$$E(G \otimes H) = \{(u_1, v_1)(u_2, v_2) \mid u_1 u_2 \in E(G) \text{ and } v_1 v_2 \in E(H) \text{ or } v_1 = v_2 \text{ and } u_1 u_2 \in E(G)\}.$$

The strong product $G \boxtimes H$ is defined as follows:
$$V(G \boxtimes H) = V(G) \times V(H),$$
$$E(G \boxtimes H) = \{(u_1, v_1)(u_2, v_2) \mid u_1 u_2 \in E(G) \text{ and } v_1 v_2 \in E(H) \text{ or } u_1 = u_2 \text{ and } v_1 v_2 \in E(H) \text{ or } v_1 = v_2 \text{ and } u_1 u_2 \in E(G)\}.$$

The lexicographic product (composition) $G[H]$ is defined as follows:
$$V(G[H]) = V(G) \times V(H),$$
$$E(G[H]) = \{(u_1, v_1)(u_2, v_2) \mid u_1 u_2 \in E(G) \text{ or } u_1 = u_2 \text{ and } v_1 v_2 \in E(H)\}.$$

The terms and concepts that we do not define can be found in [3,6,7,12].

An interval $t-$coloring [1] of a graph $G$ is a proper edge-coloring of $G$ with colors $1, 2, \ldots, t$ such that at least one edge of $G$ is colored by $i$, $i = 1, 2, \ldots, t$, and the edges incident to each vertex $v \in V(G)$ are colored by $d_G(v)$ consecutive colors.

For $t \geq 1$, let $\mathfrak{N}_t$ denote the set of graphs which have an interval $t-$coloring, and assume: $\mathfrak{N} \equiv \bigcup_{t \geq 1} \mathfrak{N}_t$. For a graph $G \in \mathfrak{N}$, the least and the greatest values of $t$ for which $G \in \mathfrak{N}_t$ are denoted by $w(G)$ and $W(G)$, respectively.

In [1,2], Asratian and Kamalian proved the following theorem.

**Theorem 1**. Let $G$ be a regular graph. Then
1. $G \in \mathfrak{N}$ if and only if $\chi'(G) = \Delta(G)$.
2. If $G \in \mathfrak{N}$ and $\Delta(G) \leq t \leq W(G)$, then $G \in \mathfrak{N}_t$.

Later, they derived some upper bounds for $W(G)$ depending on degrees and diameter of a connected graph $G \in \mathfrak{N}$.

**Theorem 2**. [2] If $G$ is a connected graph and $G \in \mathfrak{N}$, then $W(G) \leq (d(G)+1)(\Delta(G)-1)+1$. Moreover, if $G$ is also bipartite, then $W(G) \leq d(G)(\Delta(G)-1)+1$.

In [9], Petrosyan investigated interval edge-colorings of complete graphs and $n-$dimensional cubes. In particular, he proved the following two theorems.

**Theorem 3**. If $n = p2^q$, where $p$ is odd and $q$ is nonnegative, then $W(K_{2n}) \geq 4n - 2 - p - q$.

**Theorem 4**. For any $n \in \mathbb{N}$, $W(Q_n) \geq \dfrac{n(n+1)}{2}$.

In this paper interval edge-colorings of various graph products are investigated.

## 2. INTERVAL EDGE-COLORINGS OF CARTESIAN PRODUCTS OF GRAPHS

First, interval edge-colorings of Cartesian products of graphs were investigated by Giaro and Kubale in [4], where they proved the following

**Theorem 5**. If $G \in \mathfrak{N}$, then $G \square P_m \in \mathfrak{N}\,(m \in \mathbb{N})$ and $G \square C_{2n} \in \mathfrak{N}\,(n \geq 2)$.

In the same paper they proved the following theorem.

**Theorem 6**. If $G = P_{n_1} \square P_{n_2} \square \cdots \square P_{n_k}\,(n_i \in \mathbb{N})$ or $G = P_m \square C_{2n}\,(m \in \mathbb{N}, n \geq 2)$ or $G = C_{2m} \square C_{2n}\,(m, n \geq 2)$, then $G \in \mathfrak{N}$ and $w(G) = \Delta(G)$.

For $W(G)$ of these graphs, Petrosyan, Karapetyan [8] and Petrosyan, Khachatrian [11] proved the following

**Theorem 7**. If $G = P_m \square C_{2n}\,(m \in \mathbb{N}, n \geq 2)$, then $W(G) \geq 3m + n - 2$, and if $G = P_{2m} \square C_{2n}\,(m \in \mathbb{N}, n \geq 2)$, then $W(G) \geq 4m + 2n - 2$, and if $G = P_{2m} \square C_{2n+1}\,(m, n \in \mathbb{N})$, then $W(G) \geq 4m + 2n - 1$. If $G = C_{2m} \square C_{2n}\,(m, n \geq 2)$, then
$$W(G) \geq \max\{3m+n+2, 3n+m+2\},$$
and if $G = C_{2m} \square C_{2n+1}\,(m \geq 2, n \in \mathbb{N})$, then
$$W(G) \geq \begin{cases} 2m+2n+3, & \text{if } m \text{ is even,} \\ 2m+2n+2, & \text{if } m \text{ is odd.} \end{cases}$$

In [5,7], Giaro and Kubale proved the following theorem.

**Theorem 8**. If $G, H \in \mathfrak{N}$, then $G \square H \in \mathfrak{N}$ and
$$w(G \square H) \leq w(G) + w(H),\ W(G \square H) \geq W(G) + W(H).$$

We improve the lower bound in Theorem 8 for $W(G \square H)$ when $G, H \in \mathfrak{N}$ and $H$ is an $r-$regular graph. More precisely, we show that the following theorem holds.

**Theorem 9**. If $G, H \in \mathfrak{N}$ and $H$ is an $r-$regular graph, then $G \square H \in \mathfrak{N}$ and
$$W(G \square H) \geq W(G) + W(H) + r.$$

**Corollary 1**. If $G$ is an $r-$regular graph, $H$ is an $r'-$regular graph and $G, H \in \mathfrak{N}$, then $G \square H \in \mathfrak{N}$ and
$$W(G \square H) \geq W(G) + W(H) + \max\{r, r'\}.$$

**Corollary 2**. If $G_i$ is an $r_i-$regular graph, $G_i \in \mathfrak{N}$, $1 \leq i \leq n$, and $r_1 \geq r_2 \geq \cdots \geq r_n$, then $G_1 \square G_2 \square \cdots \square G_n \in \mathfrak{N}$ and
$$W(G_1 \square G_2 \square \cdots \square G_n) \geq \sum_{i=1}^n W(G_i) + \sum_{k=1}^{n-1}\sum_{i=1}^k r_i.$$

Next, we consider Hamming graphs. Recall that the Hamming graph $H(m_1, m_2, \ldots, m_n), m_i \in \mathbb{N}, 1 \leq i \leq n$, is the Cartesian product of complete graphs $K_{m_1} \square K_{m_2} \square \cdots \square K_{m_n}$. In [10], Petrosyan noted that $H(m_1, m_2, \ldots, m_n) \in \mathfrak{N}$ if and only if $m_1 m_2 \cdots m_n$ is even. Moreover, he proved the following

**Theorem 10**. If $m = p2^q$, where $p$ is odd and $q$ is nonnegative, then $H(2m, 2m, \ldots, 2m) \in \mathfrak{N}$ and
$$w(H(2m, 2m, \ldots, 2m)) = (2m-1)n,$$
$$W(H(2m, 2m, \ldots, 2m)) \geq (4m - 2 - p - q)n.$$

By Theorem 3 and Corollary 2, we can improve the lower bound in Theorem 10.

**Theorem 11**. If $m_i = p_i 2^{q_i}$, where $p_i$ is odd and $q_i$ is nonnegative, $1 \leq i \leq n$, then $H(2m_1, 2m_2, \ldots, 2m_n) \in \mathfrak{N}$ and
$$w(H(2m_1, 2m_2, \ldots, 2m_n)) = \sum_{i=1}^n (2m_i - 1),$$
$$W(H(2m_1, 2m_2, \ldots, 2m_n)) \geq \sum_{i=1}^n (4m_i - 2 - p_i - q_i) + \sum_{i=1}^{n-1} i \cdot (2m_{n-i} - 1).$$

**Corollary 3**. If $m = p2^q$, where $p$ is odd and $q$ is nonnegative, then $H(2m, 2m, \ldots, 2m) \in \mathfrak{N}$ and
$$w(H(2m, 2m, \ldots, 2m)) = (2m-1)n,$$
$$W(H(2m, 2m, \ldots, 2m)) \geq (4m - 2 - p - q) \cdot n + \dfrac{n(n-1)(2m-1)}{2}.$$

Note that Corollary 3 generalizes Theorem 4, since $H(2, 2, \ldots, 2) = Q_n$.

Also, we provide some sufficient conditions for $W(G \square H) \geq W(G) + W(H) + d(G) \cdot r$ when $G, H \in \mathfrak{N}$ and $H$ is an $r-$regular graph. In particular, we prove the following two theorems.

**Theorem 12**. If $G$ is an $r-$regular graph and $G \in \mathfrak{N}$, then $G \square C_{2n} \in \mathfrak{N}\,(n \geq 2)$ and
$$W(G \square C_{2n}) \geq W(G) + W(C_{2n}) + n \cdot r.$$

**Theorem 13**. If $G$ is an $r-$regular graph and $G \in \mathfrak{N}$, then $G \square P_m \in \mathfrak{N}\,(m \in \mathbb{N})$ and
$$W(G \square P_m) \geq W(G) + W(P_m) + (m-1) \cdot r.$$

**Corollary 4**. If $n = p2^q$, where $p$ is odd and $q$ is nonnegative, then $W(K_{2n} \square C_{2n}) \geq 2n^2 + 4n - 1 - p - q$.

**Corollary 5**. If $G$ is an $r-$regular graph and $G \in \mathfrak{N}$, then $G \square Q_n \in \mathfrak{N}\,(n \in \mathbb{N})$ and
$$W(G \square Q_n) \geq W(G) + \dfrac{n(n + 2r + 1)}{2}.$$

Note that the lower bound in Corollary 4 is close to the upper bound for $W(K_{2n} \square C_{2n})$, since $\Delta(K_{2n} \square C_{2n}) = 2n + 1$ and $d(K_{2n} \square C_{2n}) = n + 1$, by Theorem 2, we have $W(K_{2n} \square C_{2n}) \leq 2n^2 + 4n + 1$.

We also confirm the conjecture on the $n-$dimensional cube $Q_n$ [9] and show that $W(Q_n) = \dfrac{n(n+1)}{2}$ for any $n \in \mathbb{N}$.

Next, we obtain some partial results for the case when one of the factors has no interval coloring.

**Theorem 14**. For any $m \in \mathbb{N}$ and $n \in \mathbb{Z}_+$,
$$W(K_{2n+1} \square P_{2n}) \geq 2(mn+m+n) - 1,$$
$$W(K_{2n+1} \square Q_m) \geq \frac{(m+4n)(m+1)}{2}.$$

**Corollary 6.** For any $n \in \mathbb{N}$,
$$W(K_{2n+1} \square Q_n) \geq \frac{5n^2+5n}{2}.$$

Note that the lower bound in Corollary 6 is close to the upper bound for $W(K_{2n+1} \square Q_n)$, since $\Delta(K_{2n+1} \square Q_n) = 3n$ and $d(K_{2n+1} \square Q_n) = n+1$, by Theorem 2, we have $W(K_{2n+1} \square Q_n) \leq 3n^2 + 5n - 1$.

## 3. INTERVAL EDGE-COLORINGS OF TENSOR PRODUCTS OF GRAPHS

First, interval edge-colorings of tensor products of graphs were considered by Giaro and Kubale in [7], where they noted that there are $G, H \in \mathfrak{N}$ such that $G \times H \notin \mathfrak{N}$. On the other hand, Petrosyan [10] proved that if one of the factors belongs to $\mathfrak{N}$ and the other is regular, then $G \times H \in \mathfrak{N}$.

**Theorem 15.** If $G \in \mathfrak{N}$ and $H$ is an $r$ − regular graph, then $G \times H \in \mathfrak{N}$. Moreover, $w(G \times H) \leq w(G) \cdot r$ and $W(G \times H) \geq W(G) \cdot r$.

In the same paper the author formulated the following
**Problem 1.** Are there graphs $G, H \notin \mathfrak{N}$ such that $G \times H \in \mathfrak{N}$?

In [13], Yepremyan constructed graphs $G, H \notin \mathfrak{N}$ such that $G \times H \in \mathfrak{N}$. If we take the Sylvester graph $S$ as $G$ and the triangle $C_3$ as $H$, then $S \times C_3 \in \mathfrak{N}$.

## 4. INTERVAL EDGE-COLORINGS OF STRONG TENSOR PRODUCTS OF GRAPHS

First, interval edge-colorings of strong tensor products of graphs were considered by Petrosyan in [10], where he proved that if one of the factors belongs to $\mathfrak{N}$ and the other is regular, then $G \otimes H \in \mathfrak{N}$.

**Theorem 16.** If $G \in \mathfrak{N}$ and $H$ is an $r$ − regular graph, then $G \otimes H \in \mathfrak{N}$. Moreover, $w(G \otimes H) \leq w(G)(r+1)$ and $W(G \otimes H) \geq W(G)(r+1)$.

In the same paper the author formulated the following
**Problem 2.** Are there graphs $G, H \notin \mathfrak{N}$ such that $G \otimes H \in \mathfrak{N}$?

In [13], Yepremyan constructed graphs $G, H \notin \mathfrak{N}$ such that $G \otimes H \in \mathfrak{N}$. If we take the Sylvester graph $S$ as $G$ and the triangle $C_3$ as $H$, then $S \otimes C_3 \in \mathfrak{N}$.

## 5. INTERVAL EDGE-COLORINGS OF STRONG PRODUCTS OF GRAPHS

First, interval edge-colorings of strong products of graphs were considered by Giaro and Kubale [7], where they noted that there are $G, H \in \mathfrak{N}$ such that $G \boxtimes H \notin \mathfrak{N}$. On the other hand, Petrosyan [10] proved that if factors belong to $\mathfrak{N}$ and one of them is regular, then $G \boxtimes H \in \mathfrak{N}$.

**Theorem 17.** If $G, H \in \mathfrak{N}$ and $H$ is an $r$ − regular graph, then $G \boxtimes H \in \mathfrak{N}$. Moreover, $w(G \boxtimes H) \leq w(G)(r+1) + r$ and $W(G \boxtimes H) \geq W(G)(r+1) + r$.

Note that there are graphs $G$ and $H$ for which $G \boxtimes H \in \mathfrak{N}$, but $G \in \mathfrak{N}, H \notin \mathfrak{N}$. For example, $K_2 \boxtimes C_3 \in \mathfrak{N}$, but $C_3 \notin \mathfrak{N}$. Moreover, in [10], Petrosyan noted that if $G$ and $H$ are regular graphs and one of them belongs to $\mathfrak{N}$, then $G \boxtimes H \in \mathfrak{N}$. In the same paper the author formulated the following

**Problem 3.** Are there graphs $G, H \notin \mathfrak{N}$ such that $G \boxtimes H \in \mathfrak{N}$?

This problem is still open.

## 6. INTERVAL EDGE-COLORINGS OF LEXICOGRAPHIC PRODUCTS OF GRAPHS

First, interval edge-colorings of lexicographic products of graphs were considered by Giaro and Kubale in [7], where they posed the following
**Problem 4.** Does $G[H] \in \mathfrak{N}$ if $G, H \in \mathfrak{N}$?

In [10], Petrosyan proved the following two results.
**Theorem 18.** If $G \in \mathfrak{N}$, then $G[nK_1] \in \mathfrak{N}$ for any $n \in \mathbb{N}$. Moreover, $w(G[nK_1]) \leq w(G) \cdot n$ and
$$W(G[nK_1]) \geq (W(G)+1) \cdot n - 1.$$

**Theorem 19.** If $G, H \in \mathfrak{N}$ and $H$ is an $r$ − regular graph, then $G[H] \in \mathfrak{N}$. Moreover, if $|V(H)| = n$,
$w(G[H]) \leq w(G) \cdot n + r$ and $W(G[H]) \geq W(G) \cdot n + r$.

For some cases, the lower bound in Theorem 19 was improved by Yepremyan in [13].

**Theorem 20.** If $H \in \mathfrak{N}$ and $H$ is an $r$ − regular graph, then for any $n \in \mathbb{N}$, $P_n[H] \in \mathfrak{N}$ and
1. $w(P_n[H]) = \Delta(P_n[H])$,
2. $W(P_n[H]) \geq W(P_n) \cdot (|V(H)|+r) + r$.

**Theorem 21.** If $H \in \mathfrak{N}$ and $H$ is an $r$ − regular graph, then for any $n \geq 2$, $C_{2n}[H] \in \mathfrak{N}$ and
1. $w(C_{2n}[H]) = \Delta(C_{2n}[H])$,
2. $W(C_{2n}[H]) \geq (W(C_{2n})-1) \cdot (|V(H)|+r)$.

**Theorem 22.** If $H \notin \mathfrak{N}$ and $H$ is an $r$ − regular graph, then for any $n \in \mathbb{N}$, $P_{2n}[H] \in \mathfrak{N}$ and
1. $w(P_{2n}[H]) = \Delta(P_{2n}[H])$,
2. $W(P_{2n}[H]) \geq W(P_{2n}) \cdot |V(H)| + n \cdot r$.

**Theorem 23.** If $H \notin \mathfrak{N}$ and $H$ is an $r-$regular graph, then for any $n \geq 2$, $C_{2n}[H] \in \mathfrak{N}$ and

1. $w(C_{2n}[H]) = \Delta(C_{2n}[H])$,
2. $W(C_{2n}[H]) \geq (W(C_{2n}) - 1) \cdot |V(H)| + \left\lceil \dfrac{n}{2} \right\rceil \cdot r$.

Also, Yepremyan [13] noted that there are graphs $G, H \notin \mathfrak{N}$ such that $G[H] \in \mathfrak{N}$. If we take as $G$ the triangle $C_3$ and as $H$ the Petersen graph $P$, then $C_3[P] \in \mathfrak{N}$.

Finally, she proved [13] that if $T$ is a tree, then $T[P_n] \in \mathfrak{N}$ and $T[S_n] \in \mathfrak{N}$, where $S_n$ is a star $K_{1,n}$. Before we formulate these results we need some definitions.

Let $T$ be a tree and $V(T) = \{v_1, v_2, \dots, v_n\}$, $n \geq 2$. Let $P(v_i, v_j)$ be a simple path joining $v_i$ and $v_j$, $VP(v_i, v_j)$ and $EP(v_i, v_j)$ denote the sets of vertices and edges of the path, respectively.

For a simple path $P(v_i, v_j)$, define $L(v_i, v_j)$ as follows:

$L(v_i, v_j) = |EP(v_i, v_j)| + |\{uv|\ uv \in E(T), u \in VP(v_i, v_j) \setminus \{v_i, v_j\}, v \notin VP(v_i, v_j)\}|$.

Let $C(T)$ be a center of $T$, and $F(T)$ be a set of pendant vertices of $T$.

Define:
$$m(T) = \max_{u \in C(T)} \max_{v \in F(T)} L(u,v) \text{ and}$$
$$M(T) = \max_{u,v \in F(T)} L(u,v).$$

Now we can present these results.

**Theorem 24.** If $T$ is a tree, then for any $n \in \mathbb{N}$, $T[P_n] \in \mathfrak{N}$ and

1. $w(T[P_n]) \leq (m(T) + \Delta(T)) \cdot n - 1$,
2. $W(T[P_n]) \geq (M(T) + 1) \cdot n - 1$.

**Theorem 25.** If $T$ is a tree, then for any $n \in \mathbb{N}$, $T[S_n] \in \mathfrak{N}$ and

1. $w(T[S_n]) \leq (m(T) + \Delta(T)) \cdot n - 1$,
2. $W(T[S_n]) \geq (M(T) + 1) \cdot n - 1$.